\documentclass[fleqn,usenatbib]{mnras}

\usepackage{newtxtext,newtxmath}

\usepackage[T1]{fontenc}

\DeclareRobustCommand{\VAN}[3]{#2}
\let\VANthebibliography\thebibliography
\def\thebibliography{\DeclareRobustCommand{\VAN}[3]{##3}\VANthebibliography}

\usepackage{graphicx}	
\usepackage{amsmath}	

\usepackage[dvipsnames]{xcolor}

\newcommand{\orcid}[1]{\href{https://orcid.org/#1}{\includegraphics[width=10pt]{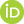}}}

\title{The metamorphosis of the Type Ib SN 2019yvr: late-time interaction}

\author[L. Ferrari et al.]{Lucía Ferrari,$^{1, 2}$\orcid{0009-0000-6303-4169}\thanks{E-mail: luciaferrari@fcaglp.unlp.edu.ar}
Gastón Folatelli,$^{1, 2, 3}$\orcid{0000-0001-5247-1486},
Hanindyo Kuncarayakti$^{4,5}$\orcid{0000-0002-1132-1366},
Maximilian Stritzinger$^{6}$\orcid{0000-0002-5571-1833},
\newauthor
Keiichi Maeda$^{7}$\orcid{0000-0003-2611-7269},
Melina Bersten$^{1, 2, 3}$,
Lili M. Román Aguilar$^{1, 2}$\orcid{0009-0000-9810-4702},
M. Manuela Sáez$^{8,9}$\orcid{0000-0001-6760-1028},
Luc Dessart$^{10}$\orcid{0000-0003-0599-8407},
\newauthor 
Peter Lundqvist$^{11}$,
Paolo Mazzali$^{12,13}$\orcid{0000-0001-6876-8284},
Takashi Nagao$^{4,14,15}$\orcid{0000-0002-3933-7861},
Chris Ashall$^{16}$\orcid{0000-0002-5221-7557}, 
Subhash Bose$^{17,6}$\orcid{0000-0003-3529-3854},
\newauthor
S. J. Brennan$^{11}$\orcid{0000-0003-1325-6235},
Y.-Z. Cai$^{18, 19, 20}$\orcid{0000-0002-7714-493X},
Rasmus Handberg$^{6}$\orcid{0000-0001-8725-4502},
Simon Holmbo$^{6}$\orcid{0000-0002-3415-322X},
Emir Karamehmetoglu$^{6}$\orcid{0000-0001-6209-838X},
\newauthor
Andrea Pastorello$^{21}$\orcid{0000-0002-7259-4624},
Andrea Reguitti$^{21, 22}$\orcid{0000-0003-4254-2724}, 
Joseph Anderson$^{23}$\orcid{0000-0003-0227-3451},
Ting-Wan Chen$^{24}$\orcid{0000-0002-1066-6098},
Lluís Galbany$^{25, 26}$\orcid{0000-0002-1296-6887}, 
\newauthor
Mariusz Gromadzki$^{27}$\orcid{0000-0002-1650-1518},
Claudia P. Guti\'errez$^{26,25}$\orcid{0000-0003-2375-2064},
Cosimo Inserra$^{28}$\orcid{0000-0002-3968-4409},
Erkki Kankare$^{4}$\orcid{0000-0001-8257-3512},
\newauthor
Tomás E. Müller Bravo$^{25,26}$\orcid{0000-0003-3939-7167},
Seppo Mattila$^{4,29}$\orcid{0000-0001-7497-2994},
Matt Nicholl$^{30}$\orcid{0000-0002-2555-3192},
Giuliano Pignata$^{31}$\orcid{0000-0003-0006-0188},
Jesper Sollerman$^{11}$\orcid{0000-0003-1546-6615},
\newauthor
Shubham Srivastav$^{32}$\orcid{0000-0003-4524-6883},
David R. Young$^{32}$\orcid{0000-0002-1229-2499}
\\
$^{1}$Instituto de Astrofísica de La Plata, CONICET, Argentina\\
$^{2}$Facultad de Ciencias Astronómicas y Geofísicas Universidad Nacional de La Plata, Paseo del Bosque S/N B1900FWA, La Plata, Argentina\\
$^{3}$Kavli Institute for the Physics and Mathematics of the Universe (WPI), The University of Tokyo, Kashiwa, 277-8583 Chiba, Japan\\
$^{4}$Department of Physics and Astronomy, University of Turku, FI-20014 Turku, Finland\\
$^{5}$Finnish Centre for Astronomy with ESO (FINCA), FI-20014 University of Turku, Finland \\
$^{6}$Department of Physics and Astronomy, Aarhus University, Ny Munkegade 120, DK-8000 Aarhus C, Denmark\\
$^{7}$Department of Astronomy, Kyoto University, Kitashirakawa-Oiwake-cho, Sakyo-ku, Kyoto, 606-8502, Japan\\
$^{8}$Interdisciplinary Theoretical and Mathematical Sciences Program (iTHEMS), RIKEN, Wako, Saitama 351-0198, Japan\\
$^{9}$Department of Physics, University of California, Berkeley, CA 94720\\
$^{10}$Institut d’Astrophysique de Paris, CNRS-Sorbonne Université, 98 bis boulevard Arago, 75014 Paris, France\\
$^{11}$The Oskar Klein Centre, Department of Astronomy, Stockholm University, AlbaNova, SE-10691 Stockholm, Sweden\\
$^{12}$Max-Planck-Institut für Astrophysik, Karl-Schwarzschild-Str. 1, D-85748 Garching, Germany\\
$^{13}$Astrophysics Research Institute, Liverpool John Moores University, 146 Brownlow Hill, Liverpool L3 5RF, UK\\
$^{14}$Aalto University Mets\"ahovi Radio Observatory, Mets\"ahovintie 114, 02540 Kylm\"al\"a, Finland\\
$^{15}$Aalto University Department of Electronics and Nanoengineering, P.O. BOX 15500, FI-00076 AALTO, Finland\\
$^{16}$Department of Physics, Virginia Tech, Blacksburg, VA 24061, USA\\
$^{17}$Department of Astronomy, The Ohio State University, 140 W. 18th Avenue, Columbus, OH 43210, USA\\
$^{18}$Yunnan Observatories, Chinese Academy of Sciences, Kunming 650216, P.R. China\\
$^{19}$Key Laboratory for the Structure and Evolution of Celestial Objects, Chinese Academy of Sciences, Kunming 650216, P.R. China\\
$^{20}$International Centre of Supernovae, Yunnan Key Laboratory, Kunming 650216, P.R. China\\
$^{21}$INAF - Osservatorio Astronomico di Padova, Vicolo dell’Osservatorio 5, 35122 Padova, Italy\\
$^{22}$INAF - Osservatorio Astronomico di Brera, Via E. Bianchi 46, 23807 Merate (LC), Italy\\
$^{23}$European Southern Observatory, Alonso de C\'ordova 3107, Casilla 19, Santiago, Chile\\
$^{24}$Graduate Institute of Astronomy, National Central University, 300 Jhongda Road, 32001 Jhongli, Taiwan\\
$^{25}$Institute of Space Sciences (ICE-CSIC), Campus UAB, Carrer de Can Magrans, s/n, E-08193 Barcelona, Spain\\
$^{26}$Institut d’Estudis Espacials de Catalunya (IEEC), E-08034 Barcelona, Spain\\
$^{27}$Astronomical Observatory, University of Warsaw, Al. Ujazdowskie 4, 00-478 Warszawa, Poland\\
$^{28}$School of Physics and Astronomy, Cardiff University, Queens Buildings, The Parade, Cardiff CF24 3AA, UK\\
$^{29}$School of Sciences, European University Cyprus, Diogenes Street, Engomi, 1516, Nicosia, Cyprus\\
$^{30}$Astrophysics Research Centre, School of Mathematics and Physics, Queens University Belfast, Belfast BT7 1NN, UK\\
$^{31}$Instituto de Alta Investigación, Universidad de Tarapacá, Casilla 7D, Arica, Chile\\
$^{32}$Astrophysics Research Centre, School of Mathematics and Physics, Queen’s University Belfast, BT7 1NN, UK
}
\date{Accepted XXX. Received YYY; in original form ZZZ}

\pubyear{2023}

\begin{document}
\label{firstpage}
\pagerange{\pageref{firstpage}--\pageref{lastpage}}
\maketitle

\begin{abstract}
We present observational evidence of late-time interaction between the ejecta of the hydrogen-poor Type Ib supernova (SN) 2019yvr and hydrogen-rich circumstellar material (CSM), similar to the Type Ib SN 2014C. A narrow H$\alpha$ emission line appears simultaneously with a break in the light-curve decline rate at around $80 - 100$~d after explosion. 
From the interaction delay and the ejecta velocity, under the assumption that the CSM is detached from the progenitor, we estimate the CSM inner radius to be located at $\sim 6.5 - 9.1 \times 10^{15}$~cm. The H$\alpha$ emission line persists throughout the nebular phase at least up to +420~d post-explosion, with a full width at half maximum (FWHM) of $\sim 2000$ km~s$^{-1}$. 
Assuming a steady mass loss, the estimated mass-loss rate from the luminosity of the H$\alpha$ line is $\sim 3 - 7 \times 10^{-5}~M_{\odot}~yr^{-1}$.
From hydrodynamical modeling and analysis of the nebular spectra, we find a progenitor He-core mass of $3 - 4 ~M_\odot$, which would imply an initial mass of $13 -15~M_\odot$.
Our result supports the case of a relatively low mass progenitor possibly in a binary system as opposed to a higher mass single star undergoing an LBV phase.

\end{abstract}
\begin{keywords}
supernovae: general -- supernovae: individual: SN~2019yvr 
--
\end{keywords}



\section{Introduction}

Supernovae (SNe) are among the most powerful explosions in the Universe, and their study provides valuable insights into a multitude of astrophysical processes including stellar evolution, the subsequent formation of compact objects, and the chemical enrichment of the Universe. A substantial fraction of SNe are associated with the collapse of the iron cores of massive stars, called core-collapse SNe (CCSNe).
CCSNe are classified based on their spectral characteristics. Objects with H features are classified as Type~II, those lacking H but exhibiting He are of Type~Ib, and those with no H nor He are of Type~Ic \citep[for a contemporary and concise review, see][and references therein]{Stritzinger2023a}. Although relatively rare, the H-poor stripped-envelope (SE) SNe are of particular interest as they are most likely linked to the death of a He or C+O star, as an analog of Wolf-Rayet stars, which experienced significant mass loss either through strong stellar winds or interaction with a binary companion
\citep[e.g.,][]{1993ApJ...411..823W,2004ApJ...612.1044P,2014ARA&A..52..487S}. Moreover, it has been proposed that the progenitor envelope could be removed by a combination of both processes, i. e. hybrid mass-loss \citep{Fang2019, Sun2023}.

A small but growing number of Type~Ib \citep[e.g.,][]{2015ApJ...815..120M,2017ApJ...837...62V,2018MNRAS.478.5050M,2020ApJ...902...55C} and Type~Ic SNe \citep{2018ApJ...854L..14K,2021A&A...650A.174T,Stritzinger2023b} exhibit signatures of circumstellar interaction (CSI) in optical wavelengths. CSI occurs when rapidly expanding SN ejecta shock a dense circumstellar material (CSM), originated from the progenitor star itself, or from a companion star \citep[e.g.,][]{Chevalier1994, 2002ApJ...572..350F,Yoon2017}.
Signatures of CSI in some cases are revealed in the spectra as narrow Balmer emission lines, reminiscent to the hallmark feature of Type IIn SNe \citep[e.g.,][]{Schlegel1990,2013A&A...555A..10T}, as well as in some cases high-ionization coronal lines, and/or excesses of flux in different regions of the electromagnetic spectrum \citep[e.g.,][]{Stritzinger2012}.

In this paper we examine SN~2019yvr, which was initially classified as a Type Ib SN \citep{2019TNSCR2736....1D}, but eventually developed SN~IIn-like features.
Based on pre-explosion Hubble Space Telescope (HST) archival images,
\cite{2021MNRAS.504.2073K} found a point source at the location of SN~2019yvr. The spectral energy distribution (SED) of the source suggests a cool and luminous progenitor candidate, in contradiction with the He star picture of a SN Ib progenitor.
They also discuss a binary scenario but also find it incompatible with a Type Ib SN progenitor. \cite{2022MNRAS.510.3701S} explored several scenarios by performing an environmental study, and proposed a binary system, composed of a hot and compact SN progenitor and a yellow hypergiant (YHG) companion. SN~2019yvr therefore provides an opportunity to understand better the pre-SN evolution of interacting events' progenitors.

The present work focuses on the appearance of the CSI features in SN 2019yvr, the late-time interaction, and particular progenitor properties such as the pre-SN mass and mass-loss rate. We will complement this analysis with an upcoming paper by Ferrari et al. (in prep.) where we will provide a detailed analysis of the full photometric and spectroscopic evolution.

\section{Observations}\label{sec:observations}
SN 2019yvr was first reported by the Asteroid Terrestrial-impact Last Alert System (ATLAS) on 2019 December 27.5 UT \citep{Tonry2019}, and classified two days later as a Type Ib SN \citep{Dimitriadis2019,Muller2019}. Based on the last non-detection from the Zwicky Transient Facility \citep[ZTF,][]{2019PASP..131a8003M}{}{}, we adopt an explosion epoch as JD $= 2458839.89 \pm 3.84$. The SN is located in the nearby Galaxy NGC 4666, which hosted the Type Ia SN ASASSN-14lp \citep{2016ApJ...826..144S}. We adopt their derived distance of $14.7 \pm 1.5$ Mpc. See Sec. 1 of the Supplementary Material for details on the discovery, the estimated explosion date and the adopted distance.

The data employed in this study are part of a larger set of multiband optical light curves and a series of spectra (Ferrari et al., in prep.). 
Here we present \textit{BVgri}-band photometry measured from images obtained by the NUTS\footnote{\url{http://nuts.sn.ie/}} (Nordic optical telescope Un-biased Transient Survey) and ASAS-SN\footnote{\url{https://www.astronomy.ohio-state.edu/asassn/}} (All-Sky Automated Survey for Supernovae) collaborations using the $2.56$-m NOT telescope equipped with ALFOSC (Andalucia Faint Object Spectrograph and Camera) and the 1-m Las Cumbres Observatory Global Telescope (LCOGT) network respectively.
The NOT images were reduced using the \texttt{pyraf}-based \texttt{ALFOSCGUI}\footnote{\url{http://graspa.oapd.inaf.it/foscgui.html}} reduction pipeline developed by E. Cappellaro, while fully processed LCOGT images were downloaded from the observatory's data archive. 

Host-galaxy subtraction was performed on all science frames using template images obtained prior to 2019.
Point-spread-function (PSF) photometry of SN~2019yvr was then measured relative to stars in the field using the Aarhus-Barcelona \texttt{FLOWS} project's automated pipeline\footnote{\url{https://github.com/SNflows}}. The photometry is tabulated in Sec. 2 of the Supplementary Material.

We also present eight low-resolution and one high-resolution optical spectra, which are summarized in the journal of spectroscopic observations (Sec. 2 of the Supplementary Material). In summary, four spectra were obtained by ePESSTO+ \citep{Smartt2015} with the European Southern Observatory (ESO) 3.58-m New Technology Telescope  \citep[NTT,][]{1984Msngr..38....9B} equipped with the ESO Faint Object Spectrograph and Camera optical (EFOSC2), three by NUTS with the NOT (+ ALFOSC), and one spectrum was obtained with the ESO 8.4-m Very Large Telescope (VLT) FOcal Reducer and low dispersion Spectrograph (FORS2, \citealt{1998Msngr..94....1A}), as part of the FORS+ Survey of Supernovae in Late Times program \citep[FOSSIL; see][]{kuncarayakti22}.
A high-resolution spectrum was obtained with the 8.2-m Subaru Telescope equipped with the High Dispersion Spectrograph \citep[HDS,][]{Noguchi2002}. The spectroscopic observations were reduced following standard techniques using the respective instrument pipelines. 

The Milky Way reddening along the line-of-sight is $E(B-V)_\textrm{MW} = 0.022$ mag \citep[i.e., $A_V^{MW} = 0.068$ mag, ][assuming $R_V=3.1$]{2011ApJ...737..103S}. After comparing the observed color curves of SN 2019yvr with the intrinsic color-curve templates from \citealt{2018A&A...609A.135S}, analyzing the diffuse interstellar band (DIB) at 5780 \AA~ and the Na I D lines in the high-resolution spectrum from the Subaru telescope, we adopt a host-galaxy reddening of $E(B-V)_{\mathrm{host}}=0.57 \pm 0.09$ mag. This value is in agreement with the estimate of \cite{2021MNRAS.504.2073K} of $0.51 ^{+0.27} _{-0.16}$ mag and with that published in \citet{Rodriguez22} of $0.56 \pm 0.09$ mag.
Details on the reddening estimation are presented in Sec. 3 of the Supplementary Material.

\section{Signatures of CSM interaction} \label{sec:CSM}

\subsection{Emergence of H\texorpdfstring{$\alpha$}{Lg} emission}\label{sec:Ha_appearance}

\begin{figure}
    \centering
    \includegraphics[width=.47\textwidth]{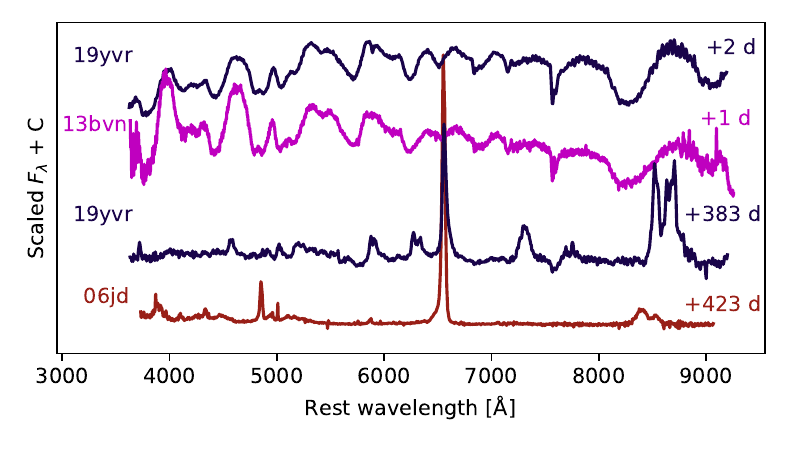}
    \caption{SN~2019yvr transitions from a SN~Ib spectrum around maximum to a Type IIn-like spectrum at late phase. Comparison data include a similar phase spectrum of the SN~Ib iPTF13bvn \citep{2014MNRAS.445.1932S} and the Type IIn SN 2006jd \citep{Stritzinger2012}.
    Spectra are corrected by redshift and not corrected by extinction.}
    \label{fig:neb_phot}
\end{figure}

SN~2019yvr showed H$\alpha$ emission at late times similarly to SN~2014C, which in that case was interpreted as a result of interaction between the ejecta and H-rich CSM \citep[e.g.][]{2015ApJ...815..120M,2017ApJ...835..140M}. In Fig.~\ref{fig:neb_phot} we compare the spectra of SN~2019yvr obtained $+$2~days (d) and $+$383~d past the epoch of $B$-band maximum, JD$=2458851.6$ (see Table 1 of the Supplementary Material). Throughout the paper we will refer the phases to this date unless otherwise specified.
While the first spectrum is consistent with that of typical SNe~Ib, the late-phase spectrum clearly exhibits the transformation to a SN~IIn-like spectrum.

\begin{figure}
    \centering
\includegraphics[width=.45\textwidth]{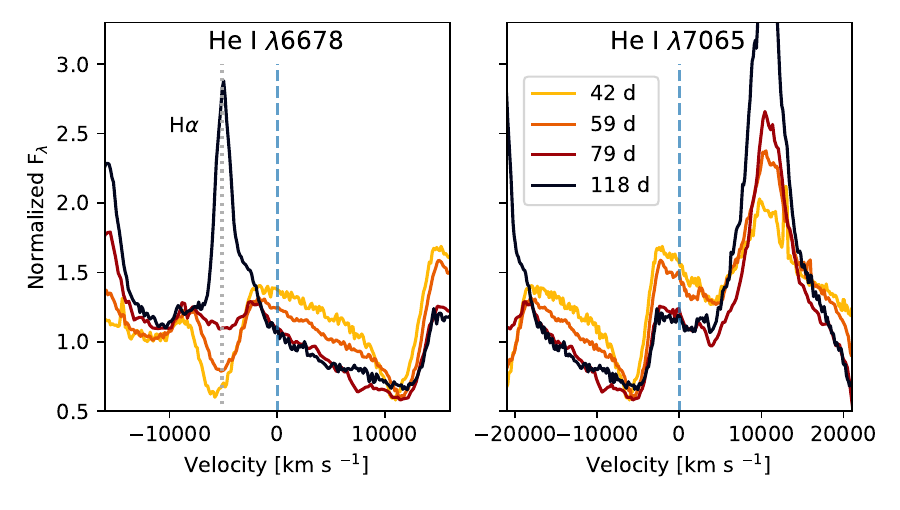}
    \caption{Evolution of the \ion{He}{i} $\lambda 6678$ and \ion{He}{i} $\lambda 7065$ features of SN~2019yvr from $+$42~d to $+$118~d plotted in velocity space. Dashed vertical lines indicate the 0 km s$^{-1}$ position, while in the left panel the dotted vertical line indicates the rest wavelength of H$\alpha$. Complete spectra are displayed in Sec. 4 of the Supplementary Material.
    }
    \label{fig:spec_evolution}
\end{figure}

The temporal coverage of SN~2019yvr observations allows us to study the moment when signatures of interaction become apparent. This is depicted in Fig.~\ref{fig:spec_evolution} where we show the evolution of the spectra around the \ion{He}{i} $\lambda6678$ (which includes H$\alpha$) and the \ion{He}{i} $\lambda7065$ lines between $+$42~d and $+$118~d. 
The +118~d spectrum clearly exhibits H$\alpha$ in emission, whereas the previous spectra show an absorption at the same wavelength. While the absorption due to \ion{He}{i} $\lambda6678$ becomes substantially weaker with time before the H$\alpha$ emission emerges, the absorption due to \ion{He}{i} $\lambda7065$ remains roughly constant.

This phenomenon can be appreciated in the evolution of the absorption pseudo-equivalent width (pEW)\footnote{The term `pseudo-equivalent width' is used because the actual continuum flux is unknown and we therefore fit a 'pseudo-continuum' between the local maxima at both sides of the line.} of both lines. 
We measured pEW using the \textit{splot} task from \texttt{IRAF} five times to account for values with their corresponding errors. For \ion{He}{i} $\lambda6678$ we obtain pEW $= 40.4 \pm 0.9$, $26.9 \pm 0.6$, and $10.4 \pm 0.3$ \AA\ at 42, 59, and 79 d, respectively, whereas for \ion{He}{i} $\lambda7065$ the values are nearly constant or even rising ($91 \pm 1.4$, $106 \pm 1.4$, and $118 \pm 2.2$ \AA). The same behavior as that of \ion{He}{i} $\lambda7065$ is seen in \ion{He}{i} $\lambda5876$ but we note that the \ion{Na}{i}~D doublet may contaminate the latter line and thus it is not shown here. We interpret the weakening of the \ion{He}{i} $\lambda6678$ absorptions as a result of the appearance of H$\alpha$ in emission at a similar wavelength.

Assuming a detached CSM from the progenitor, we conclude that the strong ejecta--CSM interaction initiated sometime prior to $+$79~d. 

\begin{figure}
   \centering
\includegraphics[width=.49\textwidth]{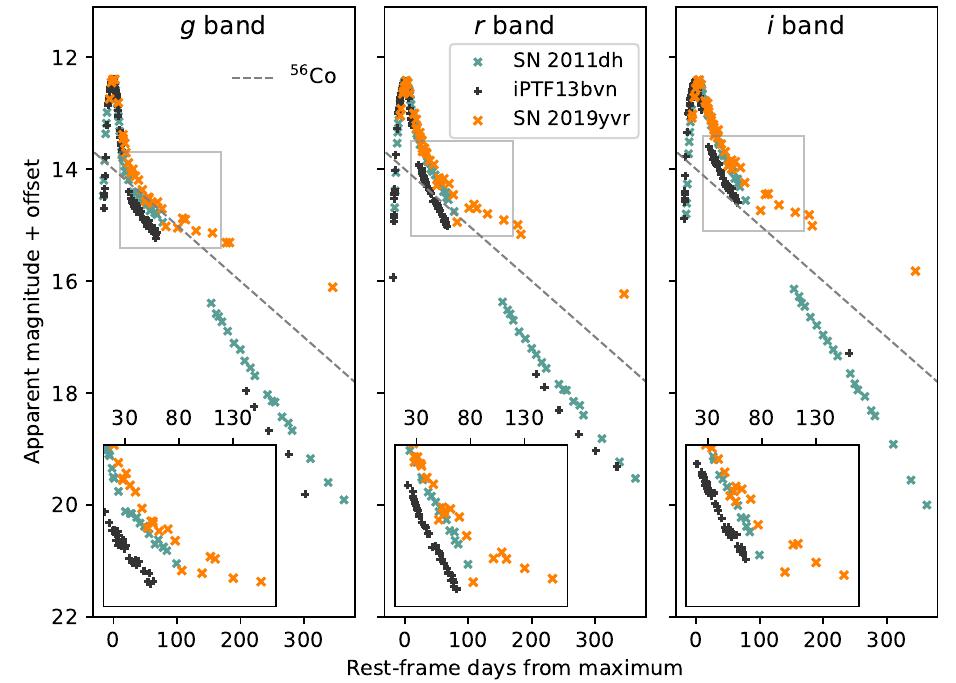}
    \caption{Light curves of SN~2019yvr compared with those of SN~2011dh \citep{2015A&A...580A.142E} and iPTF13bvn \citep{2016yCat..35930068F}, shifted in magnitudes to coincide at the peak. Gray dashed lines correspond to $^{56}$Co decay. The post-maximum light curves of SN~2019yvr begin to deviate from the normal decay line of the comparison objects beginning around $+$80~d.}
    \label{fig:LC_comparison}
\end{figure}

\subsection{Flattening in the light curves}\label{sec:LC_break}

The $gri$-band light curves are plotted in Fig. \ref{fig:LC_comparison}, compared with the Type IIb SN 2011dh and the Type Ib iPTF13bvn.
The light curves of SN~2019yvr present a characteristic break in the post-maximum decline rate, leading to a flattening after +90 d. In Fig.~\ref{fig:LC_comparison} the break is evident in all optical bands by comparison with supernovae that show a similar evolution around maximum light. Moreover, iPTF13bvn provided a very good match to the spectroscopic evolution of SN~2019yvr around maximum light.
Although the spectral match to the SN~IIb 2011dh is worse, its light curves provide a good match to our object around maximum light and are more complete than those of iPTF13bvn.

We interpret the sudden change in the slope of the light curves of SN~2019yvr as a result of an extra power source caused by sustained interaction between the SN ejecta and the CSM. This interpretation is supported by the nearly simultaneous appearance of H$\alpha$ emission in the spectra (see Section~\ref{sec:Ha_appearance}).
From close inspection of Fig.~\ref{fig:LC_comparison}, we conclude that the flattening in the light curves occurs in all bands between +70 and +90 d with respect to maximum light. 
This is in accordance with what was found in Section~\ref{sec:Ha_appearance}.

\begin{figure*}
    \centering
    \includegraphics[width=.75\textwidth]{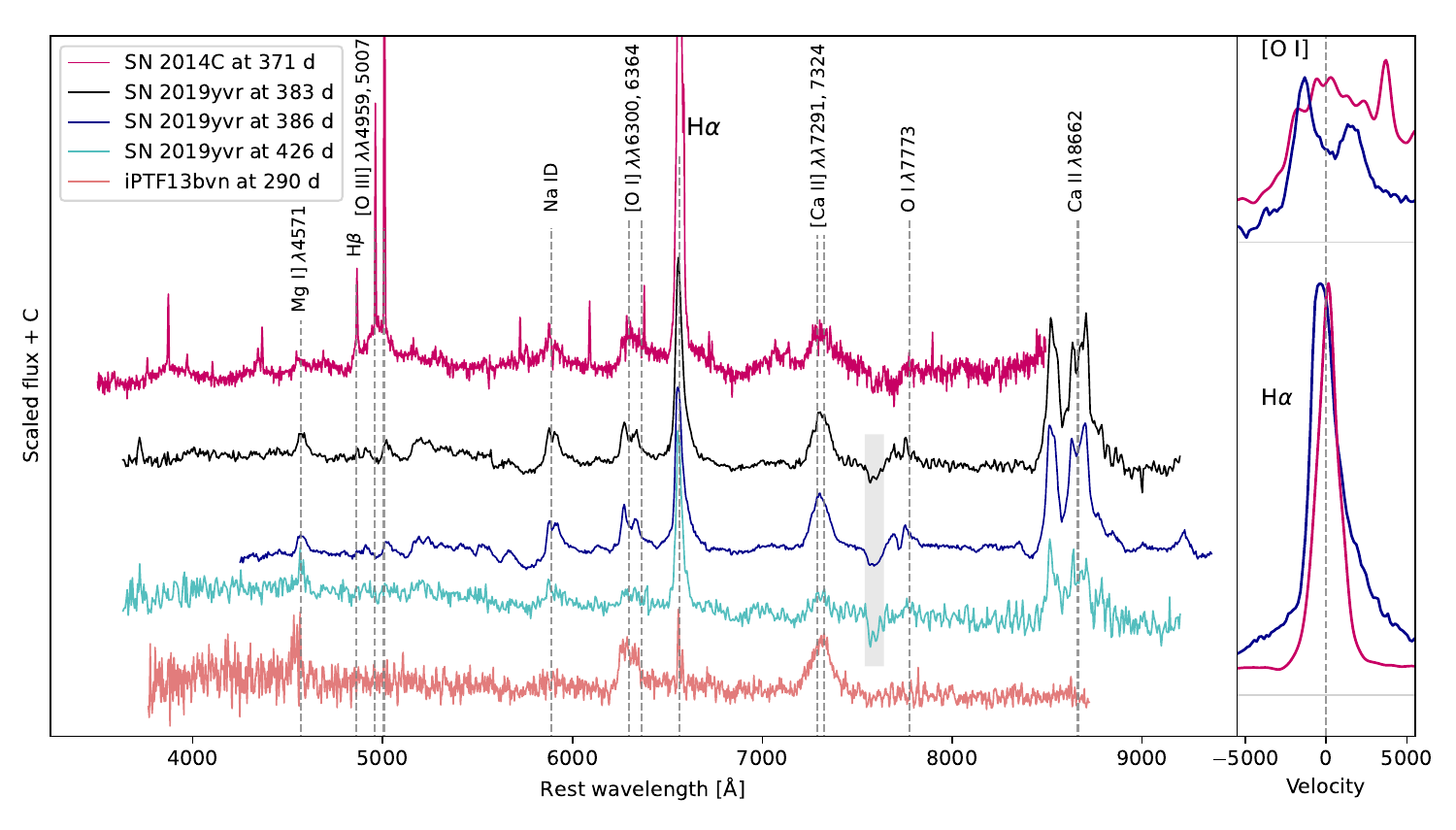}
    \caption{Left panel: nebular spectra of SN~2019yvr compared to the transitional object SN~2014C \citep{2015ApJ...815..120M}, and to its best match in the early phase, iPTF13bvn \citep{2015A&A...579A..95K}. Spectra are corrected by redshift and not corrected by extinction. Prominent emission features are identified and labeled. The telluric A-band in SN~2019yvr is marked with a gray shadow. Right panel: H$\alpha$ and oxygen doublet profiles of SN~2019yvr (+386 d, blue) and SN~2014C (+371 d, pink). The spectrum of SN~2014C has been rebinned in both panels to match the resolution of the SN~2019yvr spectrum. Both interacting events transition to a SN~IIn-like spectrum, showing strong H$\alpha$ emission. H$\alpha$ emission lines in iPTF13bvn are not associated with the SN ejecta but with an underlying \ion{H}{II} region. 
    Zero velocities are taken at 6300 and 6563 \AA.
    }
    \label{fig:nebular}
\end{figure*}

\subsection{Properties of the CSM}
Based on the light curves and spectral evolution, we have determined that the interaction power starts dominating the decay power between $+70$~d to $+90$~d (i.e., $75-105$~d post our inferred explosion epoch). 
If we assume the presence of a detached CSM structure, the interaction delay indicates a distance to its inner boundary. By adopting a maximum ejecta velocity of $\sim 10,000$~km s$^{-1}$ from the bluest extent of the \ion{He}{i} $\lambda\lambda 6678,7065$ absorption components, this gives a distance of $6.5 - 9.1 \times 10^{15}$ cm or $0.9 - 1.3 \times 10^5 R_{\odot}$. These values are comparable to those obtained for SN~2001em \citep[a SN~Ic that showed late phase H$\alpha$ in emission, ][]{2006ApJ...641.1051C}; and SN~2014C \citep{2015ApJ...815..120M}. If the CSM was expelled by stellar winds with a velocity in the range of $50 - 100$ km s$^{-1}$, the mass loss must have occurred up until $\sim 20 - 60$ years prior to the explosion (assuming a detached CSM).

We further study the properties of the CSM by analyzing the H$\alpha$ emission in the nebular phase.
We measure the H$\alpha$ luminosity, $L_{\mathrm{H}\alpha}$, to derive the mass-loss rate from \citep[e.g.][]{2018ApJ...854L..14K}
\begin{align*}
    &&\dot{M} = 2 ~\frac{L_{\mathrm{H}\alpha}}{\epsilon_{\mathrm{H}\alpha}}~\frac{v_\mathrm{wind}}{v_\mathrm{shock}^3},
\end{align*}

\noindent where $v_\mathrm{wind}$ is the wind velocity at which the material was expelled during the final stages of the star's evolution, $v_\mathrm{shock}$ is the velocity of the colliding material, and $\epsilon_{H\alpha}$ is an efficiency factor that we assume to be $0.01$ \citep{Chevalier1994}.

We scaled our +383~d and +386~d spectra so that they matched the $r$-band photometry.
They were corrected for extinction and the H$\alpha$ fluxes were then measured with \texttt{splot} in \texttt{IRAF}. We discard the measurement from the +426d spectrum because it likely suffers from host-galaxy contamination.
The resulting fluxes were $7.77 \pm 0.19 \times 10^{-14}$ erg~s$^{-1}$~cm$^{-2}$ from the $+$383~d spectrum, and $8.44 \pm 0.44 \times 10^{-14}$ erg~s$^{-1}$~cm$^{-2}$ from the +386~d spectrum, whose average yields a flux of $8.10 \pm 0.47 \times 10^{-14}$ erg~s$^{-1}$~cm$^{-2}$. With the distance given in Section~\ref{sec:observations},we obtained an H$_\alpha$ luminosity of $L_\alpha = 2.1 \pm 0.6 \times 10^{39}$ erg s$^{-1}$. For the shock velocity, we adopted 10~000 km~s$^{-1}$ based on typical values for SNe~Ib \citep{2016ApJ...827...90L}.
Assuming a wind velocity of $v_{wind} = 50 - 100$ km s$^{-1}$, the derived mass-loss rate range is $\sim 3 - 7 \times 10^{-5}~M_{\odot}~yr^{-1}$.
This is comparable to that of SN~2013df, which also shows a late-phase light curve flattening \citep{2015ApJ...807...35M}.
If a shock velocity of 2~000 km s$^{-1}$ is considered, this translates to an upper limit for the mass-loss rate of $\sim 4 - 8 \times 10^{-3}~M_{\odot}~yr^{-1}$. A high density of the CSM may prevent the appearance of [\ion{O}{iii}] lines, but would not be enough to produce electron scattering wings in H$\alpha$ (see Sec. \ref{sec:nebular}). The material could be distributed in a clumpy shell with cloud shocks of about 2~000 km s$^{-1}$ and faster, lower-density shocks in between, which would be responsible for the H$\alpha$ wings.

\section{Nebular spectra}\label{sec:nebular}

Nebular spectra of SN~2019yvr obtained on +383~d, +386~d and +426~d are plotted in Fig.~\ref{fig:nebular}, along with a +371~d spectrum of SN~2014C and a +290~d spectrum of iPTF13bvn.
These objects were classified as H-poor SNe~Ib based on their early spectroscopy\footnote{Note that some authors suggested the possible presence of H features in the spectra of iPTF13bvn and SN~2014C \citep{2015A&A...579A..95K,2015ApJ...815..120M}.}.
Similar to SN~2014C, SN~2019yvr developed a narrow $\sim 2000$~km~s$^{-1}$ H$\alpha$ emission at late times, although their profiles differ. While in SN~2019yvr the line is asymmetric and blue-shifted by $\sim 300$ km s$^{-1}$, SN~2014C showed a compound profile, with one broad $\sim 1200$ km s$^{-1}$ component overlapped with narrow $\sim 250$ km s$^{-1}$ H$\alpha$ and [\ion{N}{II}] $\lambda\lambda 6548,6583$ components \citep{2015ApJ...815..120M}. The narrow components may be linked to CSM material undergoing photoionization caused by X-rays emitted by the interaction, though these could also result from contamination by an underlying \ion{H}{ii} region. The broad component is associated with the shock or ejected material colliding with the CSM. In the right panel of Fig.~\ref{fig:nebular} we rebinned the spectrum of SN 2014C to match the resolution of the FORS spectrum of SN~2019yvr. We conclude that those narrow lines, if present, are not resolved by our observations. Another difference can be appreciated in the narrow emissions associated with H$\beta$ and [\ion{O}{iii}] $\lambda\lambda4959, 5007$ that are absent in the case of SN~2019yvr.

A striking feature is the strong \ion{Ca}{ii} near-infrared triplet, which is usually weaker than [\ion{O}{i}] $\lambda\lambda6300,6364$ and [\ion{Ca}{ii}] $\lambda\lambda7291,7324$ in SESNe \citep{2015A&A...573A..12J,2021A&A...656A..61D,2023arXiv230612092D}.
This feature is detected in the $+$282~d spectrum of SN~2014C, but with a substantially weaker intensity \citep{2015ApJ...815..120M} compared to SN~2019yvr.

Typical nebular emission lines are present, such as the aforementioned [\ion{O}{i}], [\ion{Ca}{ii}], also \ion{Na}{i}~D, and \ion{Mg}{i}] $\lambda4571$. The oxygen doublet shows a double-peaked profile, with a $\sim 1300$ km s$^{-1}$ blueshift and a FWHM of $\sim 2000$ and $\sim 2500$ km s$^{-1}$ for the bluer and redder component respectively, different from the one-component profile in SN~2014C.
The Na~I~D emission presents a broad profile, with ISM absorption on top. The [\ion{Ca}{ii}] and \ion{Mg}{i}] lines show a single component profile. Line identifications are shown in Fig.~\ref{fig:nebular}.

The H$\alpha$ profile provides insights into the geometry responsible for such emission. If the CSM was distributed in a spherical structure surrounding the SN, the profile should be box-shaped as in SN 1993J \citep{1994AJ....108.2220F,1995A&A...299..715P,Matheson2000}
and SN 2013df \citep{2015ApJ...807...35M}. 
The absence of such a profile suggests that the emission may not come from the outer layers of the ejecta interacting with the CSM, as the velocity should be higher \citep[$\sim 10000$ km s$^{-1}$, see e.g.][]{2023A&A...675A..33D}{}{}. The $\sim 2000$ km~s$^{-1}$ width the of H$\alpha$ line indicates that the slower, inner part of the ejecta are interacting with a nearby, dense CSM. This CSM may take the form of a circumstellar disk which could be produced in a binary system, although detailed modeling is required to ascertain this possibility. 
It has been proposed for SN 2014C that the H$\alpha$ emission comes from the interaction between the ejecta and a CSM with a torus-like structure \citep{Thomas2022}, which could be also the case for SN 2019yvr.

\section{Progenitor properties}\label{sec:progenitor_properties}

In this section, we aim to constrain the progenitor mass, and then link it with what was obtained in Section~\ref{sec:CSM}. For this purpose, we consider three methods: (i) the early bolometric light curve modeling, (ii) the comparison of nebular spectra with synthetic nebular spectra from \cite{2023arXiv230612092D}, and (iii) the oxygen mass estimation from the [\ion{O}{i}] $\lambda\lambda6300,6364$ flux following the procedure of \citet{2014MNRAS.439.3694J}. Our results are summarized in Fig.~\ref{fig:progenitor_mass}.

\subsection{Hydrodynamical model}\label{sec:progenitor_hydro}

We used the 1D Lagrangian hydrodynamic code \citep{2011ApJ...729...61B} to model the bolometric light curve and the photospheric velocity evolution of SN~2019yvr.
As initial configurations for our hydrodinamical models we adopted He stars of different masses from \citealt{1988PhR...163...13N}, which follow the complete evolution of the stars with ZAMS masses of 13, 15, 18, and 25 $M_{\odot}$, to the pre-SN conditions.
The simulations' free parameters are the explosion energy ($E$), the ejected mass ($\mathrm{M_{ej}}$), the mass of synthesized $^{56}$Ni ($\mathrm{M_{Ni}}$), and the extent of outward mixing of $^{56}$Ni (as a fraction of the pre-SN mass). The energy is deposited at a certain mass coordinate, $M_\textrm{cut}$, within the pre-SN structure. It is assumed that the matter inside $M_\textrm{cut}$ collapses into a compact remnant while the outer mass is ejected. 

We computed the bolometric light curve for SN~2019yvr based on $(g-r)$ and $(r-i)$ color curves using the bolometric-correction versus color calibrations for SNe~Ib given by \citet[][see their Table~2]{2014MNRAS.437.3848L}. We applied extinction and distance values as given in Section~\ref{sec:observations} to derive bolometric luminosities, and then averaged the results obtained from both color indices. Finally, to approximate photospheric velocities, we measured the Fe~{\sc ii}\,$\lambda$5169 line velocity from the location of the absorption minimum. 

Figure~\ref{fig:progenitor_mass}, panels a) and b), show the results of the modeling. Our preferred model corresponds to a pre-SN model with a mass of 3.3 $M_{\odot}$, $E=4\times 10^{50}$ erg, $M_\textrm{Ni} = 0.088~M_{\odot}$ and an extensive mixing of 0.93. We also assume a $M_\textrm{cut} = 1.5~M_\odot$, leading to an $M_\textrm{ej} = 1.8~M_{\odot}$.
However a model with a pre-SN mass of $4~M_{\odot}$ also produces a reasonable match to the data. Therefore we propose progenitors with a pre-SN mass between 3 and 4 $M_{\odot}$, which corresponds to a ZAMS mass of $13 - 15 ~M_{\odot}$.

We have also tested models with higher masses which require higher energy in order to reproduce the expansion velocities, leading to worse fitting
to the light curve (see Fig. \ref{fig:progenitor_mass}, panels a) and b)). For these more massive progenitors, we found that no set of parameters can fit both the bolometric light curve and the velocities together. Specifically, in Figure \ref{fig:progenitor_mass} we show two models corresponding to $8.0 M_{\odot}$ pre-SN models with $E = 1$ foe and $E = 5$ foe, and equal values for $M_\textrm{ej} = 6.2~M_{\odot}$ ($M_\textrm{cut} = 1.8~M_\odot$), $M_\textrm{Ni} = 0.1~M_{\odot}$, and mixing = 0.98. The first case reproduces well the velocities but not the bolometric light curve, and vice versa.

Since the lowest pre-SN mass model provided by \citet{1988PhR...163...13N} is that of 3.3 $M_\odot$, we are not able to model the bolometric light curve and photospheric velocities for lower masses and constraint the inferior limit of the progenitor star mass with this approach. The derived ZAMS mass between 13 and 15 $M_\odot$ thus corresponds to an upper limit.

\subsection{Model nebular spectra and the [\ion{O}{i}]/[\ion{Ca}{ii}] ratio}
\label{sec:progenitor_O_Ca_ratio}

The flux ratio of nebular [\ion{O}{i}] $\lambda\lambda6300,6364$ to [\ion{Ca}{ii}] $\lambda\lambda7291,7324$ lines has been suggested as an indicator of the pre-SN mass \citep[e.g.][]{1989ApJ...343..323F,maeda2007,2022ApJ...928..151F}.
We calculated these ratios on the $+$383 and the $+$426 d EFOSC2 spectra by fitting Gaussian profiles and subtracting the strong local continuum. In the case of [\ion{O}{i}], we used two Gaussians to account for the flux, as the line profile is not well fitted by one Gaussian only (see Fig \ref{fig:nebular}, right panel).
\begin{figure*}
    \centering
    \includegraphics[width=.76\textwidth]{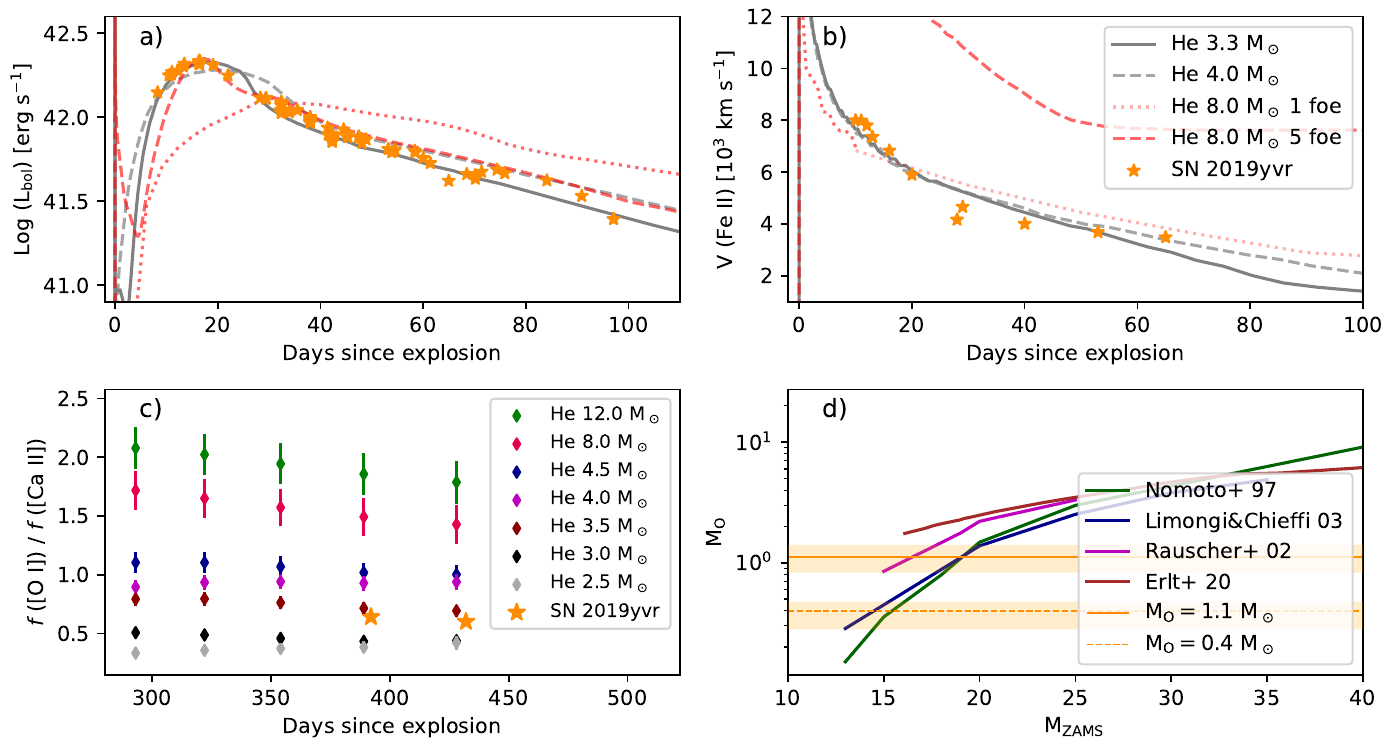}
    \caption{Progenitor mass estimation summary. Panels a) and b) show the bolometric light curve and velocity measurements of SN~2019yvr (orange stars) together with the output of the hydrodynamical models for stars with final He core masses of $3.3$, $4.0$ and $8.0~M_{\odot}$ (solid, dotted, and dashed lines, respectively; see Section \ref{sec:progenitor_hydro}). The preferred model is the least massive. In panel c) we plot the evolution of oxygen to calcium flux ratio from synthetic spectra \citep[diamonds,][]{2023arXiv230612092D} and our measurements for SN~2019yvr (orange stars), which fall in the region between 3.0 and 3.5~$M_\odot$ He core masses (see Section \ref{sec:progenitor_O_Ca_ratio}). Panel d) shows oxygen yields from \citet{1997NuPhA.616...79N}, \citet{2003ApJ...592..404L} and \citet{2002ApJ...576..323R} for different zero-age main sequence masses and the oxygen core mass derived from [\ion{O}{i}] doublet flux (see Section \ref{sec:OI_flux}).}
    \label{fig:progenitor_mass}
\end{figure*}

We use the grid of models published in \cite{2023arXiv230612092D}, where the spectral evolution between 100 and 400 d is calculated for a wide range of initial He masses. We measured the flux ratio [\ion{O}{i}]/[Ca~II] on these models by fitting a single Gaussian profile centered at $6300$ \AA~for [\ion{O}{i}] and at $7304$ \AA~for [Ca~II], since the doublets are blended. We compared these results to our measurements in SN~2019yvr at both epochs, as in Figure \ref{fig:progenitor_mass} panel {c)}. This approach yields a progenitor with a helium mass between $3.0-3.5~M_\odot$, which is in agreement with the result from the hydrodynamical model. We note, however, that the line fluxes in the model spectra ---which are computed without CSI--- are substantially smaller than those obtained for SN~2019yvr. This difference can be due to a contribution from the CSI in the line fluxes. We cannot ascertain how this may affect the flux ratios. Therefore, this caveat should be kept in mind.

\subsection{[\ion{O}{i}] doublet flux}\label{sec:OI_flux}

Following the procedure described in \cite{2014MNRAS.439.3694J}, we estimate the oxygen core minimum mass responsible for [\ion{O}{i}] doublet flux emission. 
As in Section \ref{sec:progenitor_O_Ca_ratio}, we assume that the flux comes only from the ejecta and has no contribution from the CSI. This is a strong assumption, since it has been shown that at late times the material excited by CSI can play a major role in the spectral features \citep{2023A&A...675A..33D}. Furthermore, the models by \citealt{2023arXiv230612092D} without CSI have much lower lines fluxes, as do the spectra of the SNe that do not appear to have CSI (SN 2011dh and iPFT13bvn). We thus consider this an upper limit for the pre-SN progenitor mass, and leave further analysis on how the CSI may affect the spectral features for the acompanying paper.

The flux measurement is performed as detailed in Section \ref{sec:nebular} in the dereddened spectrum at +383 d, due to its high quality and minimal host contamination.
Temperature estimation from [\ion{O}{i}] $\lambda 5577$ is not available due to the absence of the line. We therefore assume a typical temperature for these regions of $3000$ K. In these conditions, the estimated core oxygen mass is $\sim1.1 \pm 0.3~M_\odot$. If we assume a higher temperature of $3500$~K, the oxygen mass drops to $\sim 0.4 \pm 0.09~M_{\odot}$.

In both cases, following oxygen production yields from \cite{1997NuPhA.616...79N}, \cite{2002ApJ...576..323R} and \cite{2003ApJ...592..404L}, the estimates indicate a progenitor mass between $15$ and $20~M_{\odot}$ (see Fig. \ref{fig:progenitor_mass}, panel d)).
This value is somewhat higher than those obtained in Sections \ref{sec:progenitor_hydro} and \ref{sec:progenitor_O_Ca_ratio}.

\section{Conclusions}\label{sec:discussion}

We have presented light curves and spectra of SN~2019yvr that show clear signatures of late-time interaction with a CSM. Time-series observations allowed us to constrain the onset of light curve flattening and H$\alpha$ emission line. We estimated the timing of the CSI and thus the CSM distance to the progenitor, as $\sim 6.5-9.1\times10^{15}$ cm in case it is detached from the SN progenitor star. Assuming a steady $50 - 100$ km s$^{-1}$ wind velocity, this implies a mass-loss rate of $\sim 3-7 \times 10 ^{-5}~M_{\odot}~yr^{-1}$, occurring up until $\sim 20 - 60$ years prior to the explosion.

Our analysis on the progenitor mass presented in Section \ref{sec:progenitor_properties} is in contradiction with progenitors with pre-SN masses of $\ge 8~M_{\odot}$ .
Such a star may have started as a single massive star on the ZAMS and lost the outer layers via vigorous winds, but in the case of SN~2019yvr a less massive star that lost its H-rich envelope through binary interactions \citep[e.g.,][]{Fang2019,2023arXiv230700061D}{}{} is a more plausible scenario. It is also possible that the progenitor star experienced a hybrid mass-loss mechanism as that discussed by \citet{Fang2019} and \citet{Sun2023}.

The main question lies in how a progenitor with no hydrogen, as indicated by the early spectra, can lead to a H-rich SN at later times. 
\cite{2021MNRAS.504.2073K} suggested two progenitor scenarios for SN 2019yvr: a massive star that went through a series of eruptions in a luminous blue variable (LBV) phase, or a binary system that led to mass-loss episodes timed years to decades ahead of core collapse. 
\cite{2022MNRAS.510.3701S} suggested a hot and compact progenitor in a binary system with a cool and inflated YHG companion. 
Compared with these works, our results are compatible with the binary progenitor scenario and do not favor a single star going through an LBV phase.
By studying the host stellar cluster of SN 2014C, \cite{2020MNRAS.497.5118S} suggested that the progenitor could have been an $11~M_{\odot}$ star depleted by binary interaction. Otherwise, a single star should have retained its hydrogen-rich envelope to be consistent with the clusters' inferred age. If this is correct, both SNe must have gone through similar evolutionary and mass-loss paths, resulting in stripped progenitors with initial masses well below 20 $M_{\odot}$.

\section*{Data Availability}
Nebular spectra are available in the Wiserep \citep{2012PASP..124..668Y} database (\url{https://www.wiserep.org/}).

\section*{Acknowledgemets}
The Finnish National Agency for Education (EDUFI) supported this research project through an EDUFI Fellowship.
H.K. was funded by the Research Council of Finland projects 324504, 328898, and 353019.
M.D.S. is funded by the Independent Research Fund Denmark (IRFD) via Project 2 grant 10.46540/2032-00022B. 
K.M. acknowledges support from the Japan Society for the Promotion of Science (JSPS) KAKENHI grant (JP20H00174) and by the JSPS Open Partnership Bilateral Joint Research Project between Japan and Finland (JPJSBP120229923). 
NUTS is funded in part by the Instrument center for Danish Astrophysics (IDA). LCOGT network data were obtained through OPTICON (PI Stritzinger, program ID 2020A/031) and NOAO (PI Bose, program ID NOAO2020A-017) time allocations.
Based on observations made with the Nordic Optical Telescope, owned in collaboration by the University of Turku and Aarhus University, and operated jointly by Aarhus University, the University of Turku and the University of Oslo, representing Denmark, Finland and Norway, the University of Iceland and Stockholm University at the Observatorio del Roque de los Muchachos, La Palma, Spain, of the Instituto de Astrofisica de Canarias. Data were also obtained with ALFOSC, which is provided by the Instituto de Astrofisica de Andalucia (IAA) under a joint agreement with the University of Copenhagen and NOT. This work includes data collected at ESO via program IDs 1103.D-0328, 0105.D-0511 and 1106.D-0811.
We thank the Subaru staff for the data taken by the Subaru Telescope (S19B-054); the authors acknowledge the very significant cultural role and reverence that the summit of Maunakea has always had within the indigenous Hawaiian community. We are most fortunate to have the opportunity to conduct observations from this mountain.
Y.-Z. Cai is supported by the National Natural Science Foundation of China (NSFC, Grant No. 12303054) and the International Centre of Supernovae, Yunnan Key Laboratory (No. 202302AN360001).
A.P. is supported by the PRIN-INAF 2022 project "Shedding light on the nature of gap transients: from the observations to the models".
L.G. acknowledges financial support from the Spanish Ministerio de Ciencia e Innovaci\'on (MCIN), the Agencia Estatal de Investigaci\'on (AEI) 10.13039/501100011033, and the European Social Fund (ESF) "Investing in your future" under the 2019 Ram\'on y Cajal program RYC2019-027683-I and the PID2020-115253GA-I00 HOSTFLOWS project, from Centro Superior de Investigaciones Cient\'ificas (CSIC) under the PIE project 20215AT016, and the program Unidad de Excelencia Mar\'ia de Maeztu CEX2020-001058-M.
C.P.G. acknowledges financial support from the Secretary of Universities and Research (Government of Catalonia) and by the Horizon 2020 Research and Innovation Programme of the European Union under the Marie Sk\l{}odowska-Curie and the Beatriu de Pin\'os 2021 BP 00168 programme, from the Spanish Ministerio de Ciencia e Innovaci\'on (MCIN) and the
Agencia Estatal de Investigaci\'on (AEI) 10.13039/501100011033 under the PID2020-115253GA-I00 HOSTFLOWS project, and the program Unidad de Excelencia Mar\'ia de Maeztu CEX2020-001058-M.
T.E.M.B. acknowledges financial support from the Spanish Ministerio de Ciencia e Innovaci\'on (MCIN), the Agencia Estatal de Investigaci\'on (AEI) 10.13039/501100011033, and the European Union Next Generation EU/PRTR funds under the 2021 Juan de la Cierva program FJC2021-047124-I and the PID2020-115253GA-I00 HOSTFLOWS project, from Centro Superior de Investigaciones Cient\'ificas (CSIC) under the PIE project 20215AT016, and the program Unidad de Excelencia Mar\'ia de Maeztu CEX2020-001058-M.
S.M. acknowledges support from the Research Council of Finland project 350458.
M.N. is supported by the European Research Council (ERC) under the European Union’s Horizon 2020 research and innovation programme (grant agreement No.~948381) and by UK Space Agency Grant No.~ST/Y000692/1.
G.P. acknowledges support from ANID through Millennium Science Initiative Programs ICN12\_009.

\bibliographystyle{mnras}
\bibliography{bibliography}

\label{lastpage}
\end{document}


\label{firstpage}
\pagerange{\pageref{firstpage}--\pageref{lastpage}}
\maketitle

\section{Discovery, estimated explosion time and distance}\label{app:discovery}

The discovery of SN~2019yvr was reported to the Transient Name Server (TNS) by the Asteroid Terrestrial-impact Last Alert System (ATLAS) based on images obtained on 2019 December 27.5 UT (i.e., JD=2458845.02) with an apparent $o$-band AB magnitude of 15.9 \citep{Tonry2019}.
Two days after the discovery report, this transient was classified as a SN~Ib \citep{Dimitriadis2019,Muller2019}.
With J2000 coordinates of R.A. $ = 12^{h}45^{m}08^{s}.135$ and Decl. 
 $= -00^\circ27'32\farcs83$, SN~2019yvr is 
 located within the nearby spiral galaxy NGC~4666.

According to the discovery report, the last ATLAS non-detection occurred on 2019 December 11.6 UT \citep{Tonry2019}, while more constraining non-detection limits come from the Zwicky Transient Facility \citep[ZTF,][]{2019PASP..131a8003M} public data stream. According to ALeRCE \citep{Forster2021}, the best previous ZTF non-detection limit corresponds to a $g$-band magnitude of 19.36 based on an image obtained on 2019 December 18.6 UT (i.e., JD $=2458836.05$). \citet{Leonini2019} also reported to TNS a recovered detection of SN~2019yvr with a magnitude of 15.8 based on an open filter image taken on 2019 December 26.2 (i.e., JD $=2458843.73$).	
Taking the mid-point between the last ZTF non-detection and the recovered detection, we adopt the explosion epoch as JD $=2458839.89 \pm 3.84$.

NED reports a heliocentric redshift for the host galaxy NGC 4666 of $z = 0.005047$, along with a handful of Tully-Fisher based distances. Fortuitously, NGC~4666 also hosted the Type~Ia SN ASASSN-14lp that suggests a distance of $14.7\pm1.5$ Mpc, or $\mu = 30.8\pm 0.2$ mag \citep{2016ApJ...826..144S}. This value is adopted to set the absolute flux scale of SN~2019yvr. 

\section{Photometric and spectroscopic data}

Table \ref{tab:observations_detail} provides the details of the photometric observations of SN 2019yvr in the $BVgri$ bands.
Table \ref{tab:spectra_detail} gives the specifications of the spectroscopic observations included in this work.

\begin{table*}
    \centering
    \caption{Photometric observations of SN~2019yvr.}
    \label{tab:observations_detail}
    \begin{tabular}{ccccccc}
\hline
JD & Telescope+instrument   &$B$& $V$ & $g$ & $r$ &        $i$ \\
\hline  
2458847.7  & NOT+ALFOSC   &  17.21  $\pm$   0.04   &      16.10 $\pm$    0.09  &       $\cdots$       &  15.70  $\pm$   0.06  &  15.42   $\pm$    0.15  \\
2458848.8  & 1m0-09+fa04  &  17.00  $\pm$   0.08   &       $\cdots$           &  16.45  $\pm$    0.08  &  15.58  $\pm$   0.08  &  15.36   $\pm$    0.08  \\
2458851.6  & 1m0-12+fa06  &  16.69  $\pm$   0.07   &       $\cdots$           &  16.16  $\pm$    0.04  &  15.31  $\pm$   0.05  &  15.05   $\pm$    0.04  \\
2458852.7  & NOT+ALFOSC   &  16.80  $\pm$   0.07   &      15.77  $\pm$   0.05  &       $\cdots$       &  15.27  $\pm$   0.03  &  15.03   $\pm$    0.05  \\
2458853.8  & 1m0-09+fa04  &  16.71  $\pm$   0.08   &         $\cdots$        &  16.10  $\pm$    0.08  &  15.19  $\pm$   0.07  &  14.90   $\pm$    0.07  \\
2458856.9  & 1m0-06+fa07  &  16.70  $\pm$   0.09   &         $\cdots$        &  16.10  $\pm$    0.08  &  15.08  $\pm$   0.03  &  14.75   $\pm$    0.04  \\
2458856.9  & 1m0-08+fa05  &  16.73  $\pm$   0.30   &         $\cdots$        &       $\cdots$       &  15.16  $\pm$   0.07  &  14.83   $\pm$    0.07  \\
2458859.6  & 1m0-10+fa16  &  16.93  $\pm$   0.15   &      15.63  $\pm$     0.08 &       $\cdots$       &  15.11  $\pm$   0.06  &  14.74   $\pm$    0.06  \\
2458862.5  & 1m0-10+fa16  &  17.19  $\pm$   0.44   &          $\cdots$       &  16.52  $\pm$    0.27  &  15.30  $\pm$   0.11  &  14.83   $\pm$    0.09  \\
2458868.8  & 1m0-09+fa04  &  17.96  $\pm$   0.12   &         $\cdots$        &       $\cdots$       &  15.67  $\pm$   0.06  &  15.12   $\pm$    0.06  \\
2458869.9  & 1m0-06+fa07  &  18.12  $\pm$   0.20   &         $\cdots$        &  17.16  $\pm$    0.08  &  15.66  $\pm$   0.05  &  15.15   $\pm$    0.06  \\
2458869.9  & 1m0-08+fa05  &  18.17  $\pm$   0.19   &         $\cdots$        &  17.09  $\pm$    0.14  &     $\cdots$        &     $\cdots$          \\
2458872.9  & 1m0-04+fa03  &  18.29  $\pm$   0.29   &     16.62  $\pm$     0.08   &       $\cdots$       &  15.94  $\pm$   0.05  &  15.16   $\pm$    0.05  \\
2458872.9  & 1m0-08+fa05  &  18.57  $\pm$   0.28   &       $\cdots$          &  17.24  $\pm$    0.10  &  15.77  $\pm$   0.07  &  15.25   $\pm$    0.07  \\
2458874.4  & 1m0-12+fa06  &  17.76  $\pm$   1.43   &       $\cdots$          &  17.41  $\pm$    0.14  &  15.92  $\pm$   0.07  &  15.31   $\pm$    0.08  \\
2458878.4  & 1m0-10+fa16  &  18.55  $\pm$   0.26   &       $\cdots$          &  17.59  $\pm$    0.11  &  16.01  $\pm$   0.06  &  15.41   $\pm$    0.08  \\
2458878.5  & 1m0-13+fa14  &  18.41  $\pm$   0.35   &     16.75  $\pm$   0.11    &     $\cdots$         &  16.08  $\pm$   0.06  &  15.47   $\pm$    0.07  \\
2458881.9  & 1m0-06+fa07  &  19.05  $\pm$   0.51   &     16.84  $\pm$    0.15    &     $\cdots$         &  16.15  $\pm$   0.07  &  15.54   $\pm$    0.07  \\
2458882.8  & 1m0-05+fa15  &  18.76  $\pm$   0.13   &        $\cdots$         &  17.78  $\pm$    0.08  &  16.29  $\pm$   0.05  &  15.59   $\pm$    0.04  \\
2458882.8  & 1m0-09+fa04  &  18.81  $\pm$   0.13   &        $\cdots$         &  17.76  $\pm$    0.09  &  16.34  $\pm$   0.08  &  15.74   $\pm$    0.10  \\
2458885.7  & 1m0-04+fa03  &  18.58  $\pm$   0.32   &        $\cdots$         &  17.71  $\pm$    0.19  &  16.29  $\pm$   0.05  &  15.66   $\pm$    0.04  \\
2458888.6  & 1m0-12+fa06  &  18.72  $\pm$   0.47   &     16.96  $\pm$   0.11    &     $\cdots$         &  16.36  $\pm$   0.07  &  15.77   $\pm$    0.06  \\
2458889.4  & 1m0-13+fa14  &  18.93  $\pm$   1.01   &         $\cdots$        &  17.83  $\pm$    0.42  &  16.38  $\pm$   0.13  &  15.63   $\pm$    0.09  \\
2458894.4  & 1m0-12+fa06  &  18.76  $\pm$   0.34   &         $\cdots$        &  17.90  $\pm$    0.22  &  16.50  $\pm$   0.07  &  15.89   $\pm$    0.06  \\
2458893.7  & NOT+ALFOSC   &  18.96  $\pm$   0.09   &    17.29  $\pm$    0.07    &     $\cdots$         &  16.51  $\pm$   0.06  &       $\cdots$        \\
2458900.5  & 1m0-13+fa14  &  18.84  $\pm$   0.20   &        $\cdots$         &  18.07  $\pm$    0.13  &  16.57  $\pm$   0.06  &  16.03   $\pm$    0.06  \\
2458905.5  & 1m0-12+fa06  &  19.24  $\pm$   0.16   &        $\cdots$         &  18.28  $\pm$    0.08  &  16.95  $\pm$   0.06  &  16.27   $\pm$    0.07  \\
2458909.0  & 1m0-11+fa12  &  19.11  $\pm$   0.32   &        $\cdots$         &  18.21  $\pm$    0.18  &  16.82  $\pm$   0.06  &  16.23   $\pm$    0.06  \\
2458910.7  & 1m0-09+fa04  &  19.08  $\pm$   0.16   &     17.69 $\pm$    0.10   &  18.21  $\pm$    0.10  &  16.85  $\pm$   0.09  &  16.33   $\pm$    0.11  \\
2458910.6  & NOT+ALFOSC   &  19.17  $\pm$   0.15   &     17.65 $\pm$    0.11   &  18.24  $\pm$    0.09  &  16.88  $\pm$   0.11  &  16.17   $\pm$    0.07  \\
2458916.4  & 1m0-13+fa14  &  19.03  $\pm$   0.89   &        $\cdots$         &  18.31  $\pm$    0.56  &  16.83  $\pm$   0.18  &  16.20   $\pm$    0.11  \\
2458924.6  & 1m0-05+fa15  &  19.01  $\pm$   0.18   &        $\cdots$         &  18.29  $\pm$    0.09  &  16.92  $\pm$   0.07  &  16.31   $\pm$    0.07  \\
2458931.4  & 1m0-13+fa14  &  19.23  $\pm$   0.25   &        $\cdots$         &  18.41  $\pm$    0.16  &  17.12  $\pm$   0.07  &  16.58   $\pm$    0.07  \\
2458937.6  & NOT+ALFOSC   &  19.69  $\pm$   0.13   &   17.95  $\pm$  0.07      &  18.73  $\pm$    0.07  &  17.60  $\pm$   0.04  &       $\cdots$        \\
2458956.5  & NOT+ALFOSC   &  19.79  $\pm$   0.10   &   18.45  $\pm$  0.08      &  18.75  $\pm$    0.09  &  17.35  $\pm$   0.04  &  17.08   $\pm$    0.08  \\
2458964.2  & 1m0-03+fa19  &  19.31  $\pm$   0.31   &   17.90  $\pm$  0.15      &  18.58  $\pm$    0.23  &  17.29  $\pm$   0.11  &  16.79   $\pm$    0.09  \\
2458968.6  & NOT+ALFOSC   &  19.69  $\pm$   0.31   &   18.12  $\pm$  0.12      &  18.60  $\pm$    0.10  &  17.36  $\pm$   0.19  &  16.78   $\pm$    0.10  \\
2458985.5  & NOT+ALFOSC   &  19.88  $\pm$   0.16   &   18.56  $\pm$  0.07      &  18.81  $\pm$    0.07  &  17.46  $\pm$   0.05  &  16.97   $\pm$    0.06  \\
2459011.4  & NOT+ALFOSC   &  19.76  $\pm$   0.32   &   18.28  $\pm$  0.13      &  18.84  $\pm$    0.10  &  17.57  $\pm$   0.05  &  17.11   $\pm$    0.06  \\
2459033.4  & NOT+ALFOSC   &       $\cdots$       &   18.45  $\pm$  0.16      &  19.02  $\pm$    0.14  &  17.65  $\pm$   0.06  &  17.15   $\pm$    0.10  \\
2459038.4  & NOT+ALFOSC   &       $\cdots$       &   18.61  $\pm$  0.09      &  19.01  $\pm$    0.07  &  17.82  $\pm$   0.03  &  17.35   $\pm$    0.06  \\
2459201.7  & NOT+ALFOSC   &       $\cdots$       &   19.97  $\pm$  0.25      &  19.81  $\pm$    0.18  &  18.88  $\pm$   0.16  &  18.16   $\pm$    0.43  \\
\hline
    \end{tabular}
\end{table*}

\begin{table*}
    \centering
    \caption{Journal of spectroscopic observations of SN~2019yvr. The uncertainties were computed by adding in quadrature the photometry error and the nightly zero-point error.}
    \label{tab:spectra_detail}
    \begin{tabular}{cccc}
        \hline
        JD &Phase [d]$^{a}$& Telescope+instrument  & $\mathrm{t_{exp}}$ [s] \\
        \hline 
        2458853.0   & +1    & Subaru+HDS   & 5400  \\
        2458853.8   & +2    & NTT+EFOSC2   &  180  \\
        2458893.6   & +42   & NOT+ALFOSC   &  3600  \\
        2458910.6   & +59   & NOT+ALFOSC   &  3600  \\
        2458930.7   &  +79  & NTT+EFOSC2   &  1500  \\
        2458969.4   & +118  & NOT+ALFOSC   &  2400  \\
        2459234.7   &  +383 & NTT+EFOSC2   &  3200  \\
        2459237.8   &  +386 & VLT+FORS2    &  2740  \\
        2459277.8   &  +426 &  NTT+EFOSC2  &  3100  \\
        \hline
    \end{tabular}\\
   {$^a$Phase is relative to the epoch of $B$-band maximum, i. e., JD $=2458851.6$.}
\end{table*}

\section{Reddening estimation}\label{app:reddening}

Details on the reddening estimation methods are presented below. We have considered the typical value of $R_V = 3.1$. 
\citet{2021MNRAS.504.2073K} derive an $R_V$ of 4.7$^{+1.3}_{-3}$, but we consider it to be unconstrained. This is most probably due to the lack of near-infrared colors \citep[see][]{Folatelli2010}. Our assumption is well within 1 $\sigma$ of the value derived by \citet{2021MNRAS.504.2073K}.

\subsection{Color curves}\label{app:color_curves}

\begin{figure}\label{fig:color_curves}
    \centering
    \includegraphics[width=.47\textwidth]{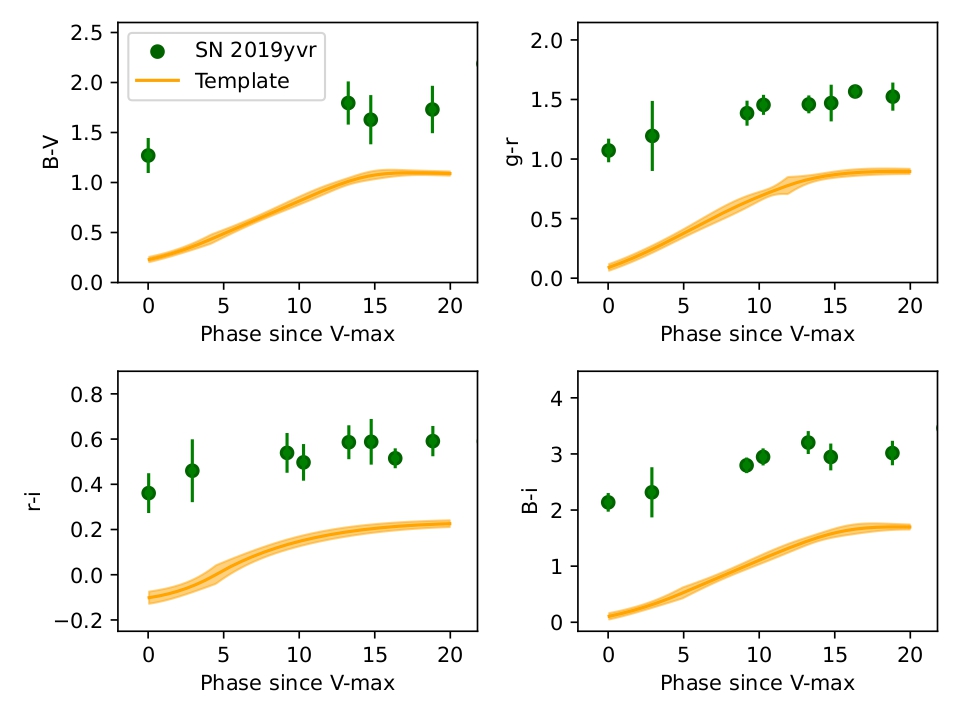}
    \caption{Apparent $(B-V)$, $(g-r)$, $(r-i)$, and $(B-i)$ color-curves of SN~2019yvr (green dots) extending between $+$0~d to $+$20~d, compared with the intrinsic color-curve templates from \citealt{2018A&A...609A.135S} (orange lines). 
    }
    \label{fig:enter-label}
\end{figure}

We estimated the host-galaxy reddening from color curves for $(B-V)$, $(g-r)$, $(r-i)$ and $(B-i)$ comparing them with templates from \citealt{2018A&A...609A.135S} (see Figure \ref{fig:color_curves}). 
We calculated the color excess for each point of the color curve and then computed an average color excess for each one of the four color indices. Finally, we transformed them to $E(B-V)$ assuming an extinction law of \citet{1989ApJ...345..245C} with $R_V=3.1$ \citep[][Table 6]{2011ApJ...737..103S} and took a weighted average. The host-galaxy reddening results $E(B-V)_\mathrm{host}=0.57 \pm 0.09$ mag.

\subsection{DIB and \ion{Na}{i} D absorption}
\begin{figure}\label{fig:subaru}
    \centering
    \includegraphics[width=.47\textwidth]{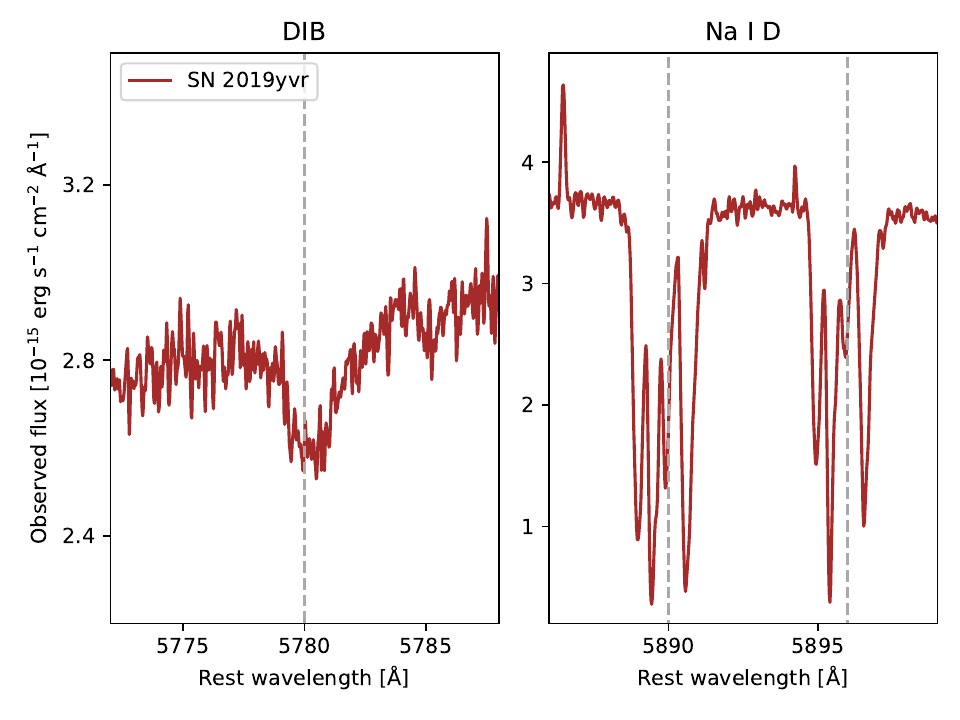}
    \caption{Subaru+HDS spectrum of SN 2019yvr obtained on $+2$~d (i. e., 2020 January 4.6 UT). Left panel: the diffuse interstellar band (DIB) at 5780 \AA. Right panel: the \ion{Na}{i}~D interstellar lines. 
    Gray dashed lines indicate the rest wavelength for all features.}
    \label{fig:DIBs}
\end{figure}

The equivalent width (EW) of absorption features of the diffuse interstellar band (DIB) at 5780 \AA~and \ion{Na}{i} D have been used for reddening estimation \citep{2012MNRAS.426.1465P,2013ApJ...779...38P}. These features are present in our high-resolution spectrum from the Subaru telescope (see Figure \ref{fig:subaru}). The DIB absorption feature is found to have an EW $\sim 0.26$~\AA, which leads to a reddening of $E(B-V)_\mathrm{host} = 0.4$ mag through the method of \citealt{2013ApJ...779...38P}, with an uncertainty of $\sim 50\%$. In the case of \ion{Na}{i}~D, its EW is $\sim 2.1$~\AA, falling at the higher end of the relation proposed by \citet{2012MNRAS.426.1465P} and thus not considered reliable for reddening estimation. Its profile with absorptions at four different velocities is an indicator that multiple clouds are located between the SN and the observer. 

We find the method based on the color curves (see Section \ref{app:color_curves}) to be more reliable, and consistent with values derived from the DIB absorption, and therefore we adopt the $E(B-V)_\mathrm{host}$ value obtained in Section \ref{app:color_curves} for our analysis.

\section{SN 2019yvr spectral series}\label{app:full_spectra}

\begin{figure}
    \centering
    \includegraphics[width=.5\textwidth]{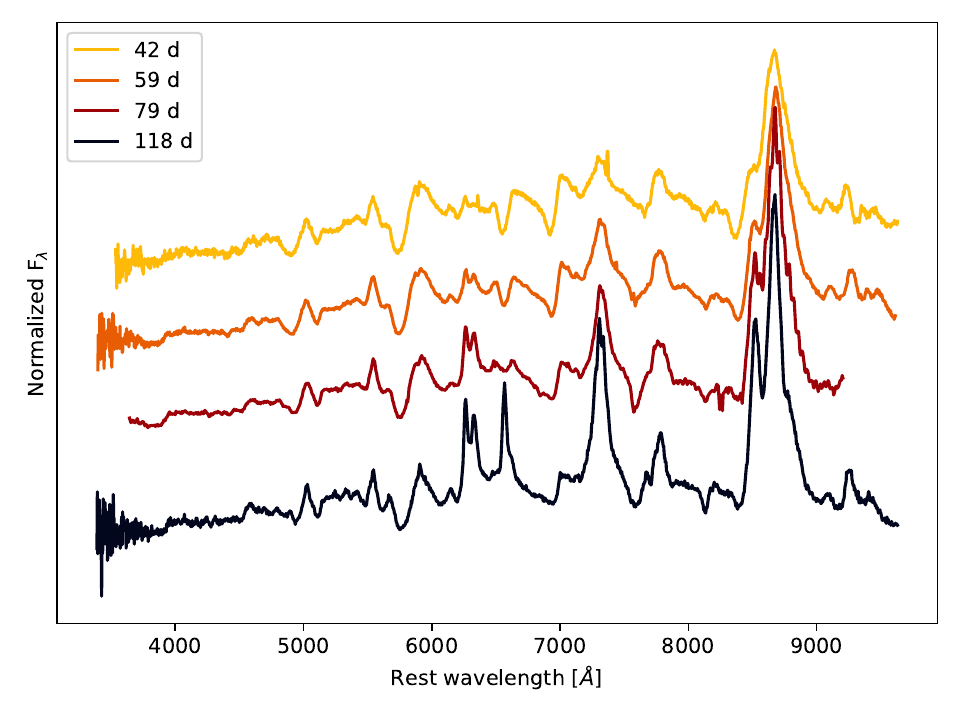}
    \caption{Optical low-resolution spectra of SN~2019yvr spanning +42~d and +118~d.}
    \label{fig:full_spectra}
\end{figure}

In Section 3.1, four spectra are presented showing the regions of interest for our analysis, showing the He I $\lambda 6678$ and He I $\lambda 7065$ lines. In Figure \ref{fig:full_spectra} these spectra are displayed entirely. The evolution spans from +42 to +118 d post-maximum and depicts the rise of the H$\alpha$ emission line. Observations details can be found in Table \ref{tab:spectra_detail}.

\bibliographystyle{mnras}
\bibliography{bibliography}

\label{lastpage}